\documentclass[
superscriptaddress,
 amsmath,amssymb,
 aps,
reprint
]{revtex4-2}

\usepackage[utf8]{inputenc}
\usepackage{color}
\usepackage{graphicx}
\usepackage{dcolumn}
\usepackage{bm}
\usepackage{epstopdf}
\usepackage{gensymb}
\usepackage{hyperref}
\usepackage{graphicx}
\usepackage{hyperref}
\usepackage{caption}  

\usepackage[compatibility=false]{caption}
\usepackage{ragged2e}

\usepackage{float}

\begin{document}
\raggedbottom

\title{Mechanical control of competing magnetic order in crystalline MnPtGa membranes}

\author{Rahul S. Rawat}
\affiliation{Materials Science and Engineering, University of Wisconsin-Madison, Madison, WI 53706, United States of America}

\author{Fan Fei}
\affiliation{Materials Science and Engineering, University of Wisconsin-Madison, Madison, WI 53706, United States of America}

\author{Zachary T. LaDuca}
\affiliation{Materials Science and Engineering, University of Wisconsin-Madison, Madison, WI 53706, United States of America}

\author{Tamalika Samanta}
\affiliation{Materials Science and Engineering, University of Wisconsin-Madison, Madison, WI 53706, United States of America}

\author{Katherine Su}
\affiliation{Materials Science and Engineering, University of Wisconsin-Madison, Madison, WI 53706, United States of America}

\author{Michael S. Arnold}
\affiliation{Materials Science and Engineering, University of Wisconsin-Madison, Madison, WI 53706, United States of America}

\author{Jun Xiao}
\affiliation{Materials Science and Engineering, University of Wisconsin-Madison, Madison, WI 53706, United States of America}

\author{Jason K. Kawasaki} \email{Author to whom correspondence should be addressed: Jason Kawasaki, jkawasaki@wisc.edu}
\affiliation{Materials Science and Engineering, University of Wisconsin-Madison, Madison, WI 53706, United States of America}

\date{\today}

\begin{abstract}

MnPtGa hosts competing magnetic states, including ferromagnetic, canted antiferromagnetic, and spin-density-wave (SDW) order in the centrosymmetric $P6_3/mmc$ structure, while a related inversion-broken structure supports chiral skyrmions. Controlling this competition motivates materials platforms that enable tunable strain and symmetry breaking, together with probes of SDW order compatible with ultrathin samples. Here, we demonstrate single-crystalline MnPtGa membranes grown by molecular beam epitaxy on graphene/Ge(111) and released by mechanical exfoliation. X-ray and electron diffraction confirm high crystalline quality. SQUID magnetometry reveals a 140 K anomaly in the zero-field-cooled $dM/dT$ that persists after exfoliation, while time-resolved reflectivity shows a coincident peak in the electronic relaxation time consistent with a quasiparticle–phonon bottleneck associated with a putative SDW gap. Intentional rippling suppresses the $\sim 140$ K magnetic anomaly, demonstrating mechanical control of the low-temperature state. These results establish MnPtGa membranes as a platform for detecting and strain-tuning competing magnetic orders.

\end{abstract}

\maketitle

\section{Introduction} 

Competing exchange interactions can stabilize multiple magnetic states with comparable energies, making magnetic order highly sensitive to lattice symmetry and structural distortion. Hexagonal MnPtGa is a particularly rich example. In the centrosymmetric $P6_3/mmc$ structure, bulk neutron diffraction reveals a sequence of ferromagnetic (FM), canted-antiferromagnetic (CAF), and incommensurate spin-density-wave (SDW) order upon cooling, together with pronounced magnetostructural coupling \cite{cooley2020evolution}. Recent first-principles calculations likewise find ferromagnetic, antiferromagnetic, canted, and spiral configurations separated by relatively small energy scales \cite{fecher2025magnetic}. In contrast, an inversion-broken trigonal $P3m1$ polymorph is reported to host Néel skyrmions \cite{srivastava2020observation}, although the origin of this symmetry lowering relative to the more commonly observed $P6_3/mmc$ structure remains unclear. These results point to a shallow magnetic energy landscape in which modest structural perturbations can qualitatively alter the magnetic ground state.

Continuous control of this competing-order landscape remains challenging. Conventional epitaxial strain can modify magnetic interactions, but is largely fixed by the substrate and growth conditions. Freestanding crystalline membranes instead permit continuously variable strain and bending-induced strain gradients \cite{du2021epitaxy,du2023strain,laduca2024cold}, while growth across a weakly bonded graphene interface may also reduce substrate clamping prior to release. These features offer potential routes to tune exchange competition, modify the SDW modulation, and access symmetry-lowered magnetic states. 

A complementary challenge is identifying modulated magnetic order in small-volume samples. In MnPtGa, the bulk SDW has been established most directly by neutron diffraction \cite{cooley2020evolution}, but the weak scattering volume of ultrathin (few nm) films and membranes makes such measurements difficult. This distinction is particularly relevant because previous 60 nm epitaxial MnPtGa films on sapphire exhibited a similar low-temperature magnetic anomaly that neutron diffraction identified as a commensurate canted state without detectable SDW order \cite{ibarra2022noncollinear}. Optical probes sensitive to changes in electronic relaxation could therefore provide complementary evidence for distinguishing these competing low-temperature states.

Here, we demonstrate single-crystalline MnPtGa membranes grown by molecular beam epitaxy (MBE) on graphene/Ge(111) and released by mechanical exfoliation. The supported films exhibit high crystalline quality and lattice parameters close to the bulk values. Temperature-dependent magnetometry of both the supported film and membrane reproduces key qualitative features of bulk MnPtGa, including a low-temperature transition near 140 K. Time-resolved reflectivity of the supported MnPtGa/graphene/Ge(111) heterostructure reveals a coincident peak in the electronic relaxation time, consistent with a relaxation bottleneck associated with density-wave formation. The 140 K transition persists after release but is suppressed by intentional rippling, demonstrating mechanical control of the low-temperature magnetic state. Together, these results support a putative SDW-containing state and establish MnPtGa membranes as a platform for mechanically tuning competing magnetic orders.

\section{Results}

\begin{figure}[t]
    \centering
    \includegraphics[width=1.03
\columnwidth]{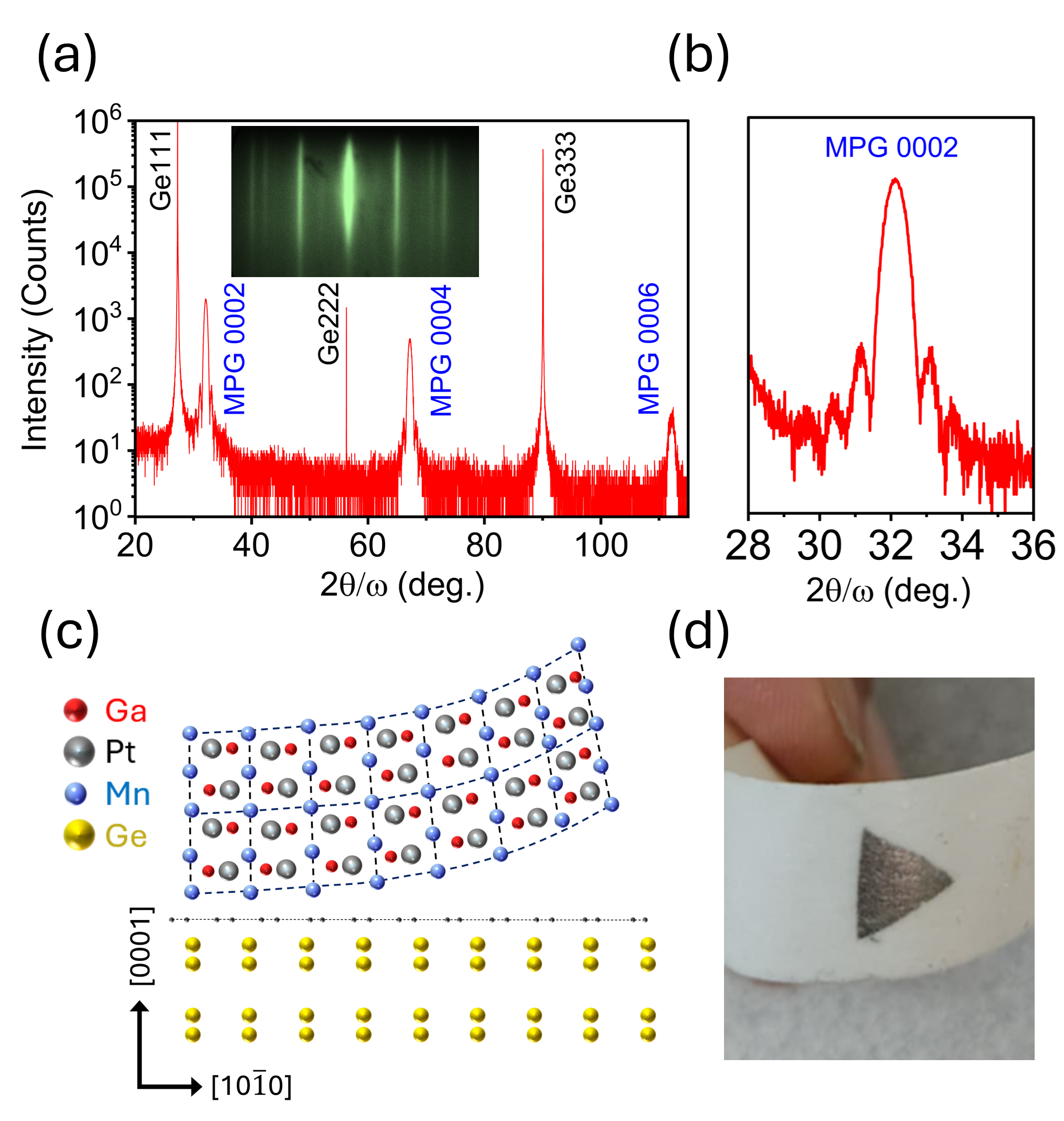}
    \caption{
    \textbf{Synthesis and exfoliation of MnPtGa membrane from graphene/Ge(111).}
    (a) Out-of-plane $2\theta-\omega$ x-ray diffraction (Cu $K\alpha$) of MnPtGa grown on graphene/Ge(111) showing $000L$ reflection. Inset: RHEED pattern along the Ge $\langle110\rangle$ azimuth. (b) Out-of-plane $2\theta-\omega$ of $0002$ MnPtGa. The inset shows Laue fringes around the $0002$ reflection, indicating smooth and uniform film growth. (c) Crystal structure of MnPtGa exfoliated from graphene/Ge (111). (d) Photo of exfoliated MnPtGa membrane.}
    \label{fig:structure}
\end{figure}

Fig. \ref{fig:structure} demonstrates epitaxial synthesis and mechanical exfoliation of MnPtGa membranes. MnPtGa films are grown by MBE on CVD-grown \cite{Kiraly2015_GrGe} epitaxial graphene/Ge(111) using a room temperature seed and anneal approach to suppress dewetting, following Ref.\cite{laduca2024cold} (Methods). After a total thickness of 11 nm, the resulting film displays a streaky reflection high energy electron diffraction pattern indicative of a smooth epitaxial film surface (Fig. \ref{fig:structure}(a) inset).
Symmetric $\omega-2\theta$ x-ray diffraction reveals an epitaxial MnPtGa film with only the expected even-index $000L$ reflections, consistent with the inversion symmetric $P6_3/mmc$ structure. Sharp Laue fringes indicate a smooth film with sharp crystalline interfaces (Fig \ref{fig:structure}a,b). 
The measured lattice constants of $c=5.572$ {\AA} and $a=4.333$ {\AA}, extracted from the $\omega-2\theta$ and asymmetric reciprocal space maps [SI Fig.~2(b)--(d)], are closer to the bulk lattice parameters ($a=4.3304$ \AA, $c=5.5726$ \AA) \cite{cooley2020evolution} than previous thicker epitaxial films grown on c-plane sapphire \cite{ibarra2022noncollinear}, despite the smaller 11 nm thickness. This may reflect differences in stoichiometry and/or reduced epitaxial clamping across the weakly bonded graphene interface. The resulting mechanical decoupling from the Ge substrate enables exfoliation using adhesive tape (Fig. \ref{fig:structure}d).

\begin{figure*}
    \centering
\includegraphics[width=1\linewidth]{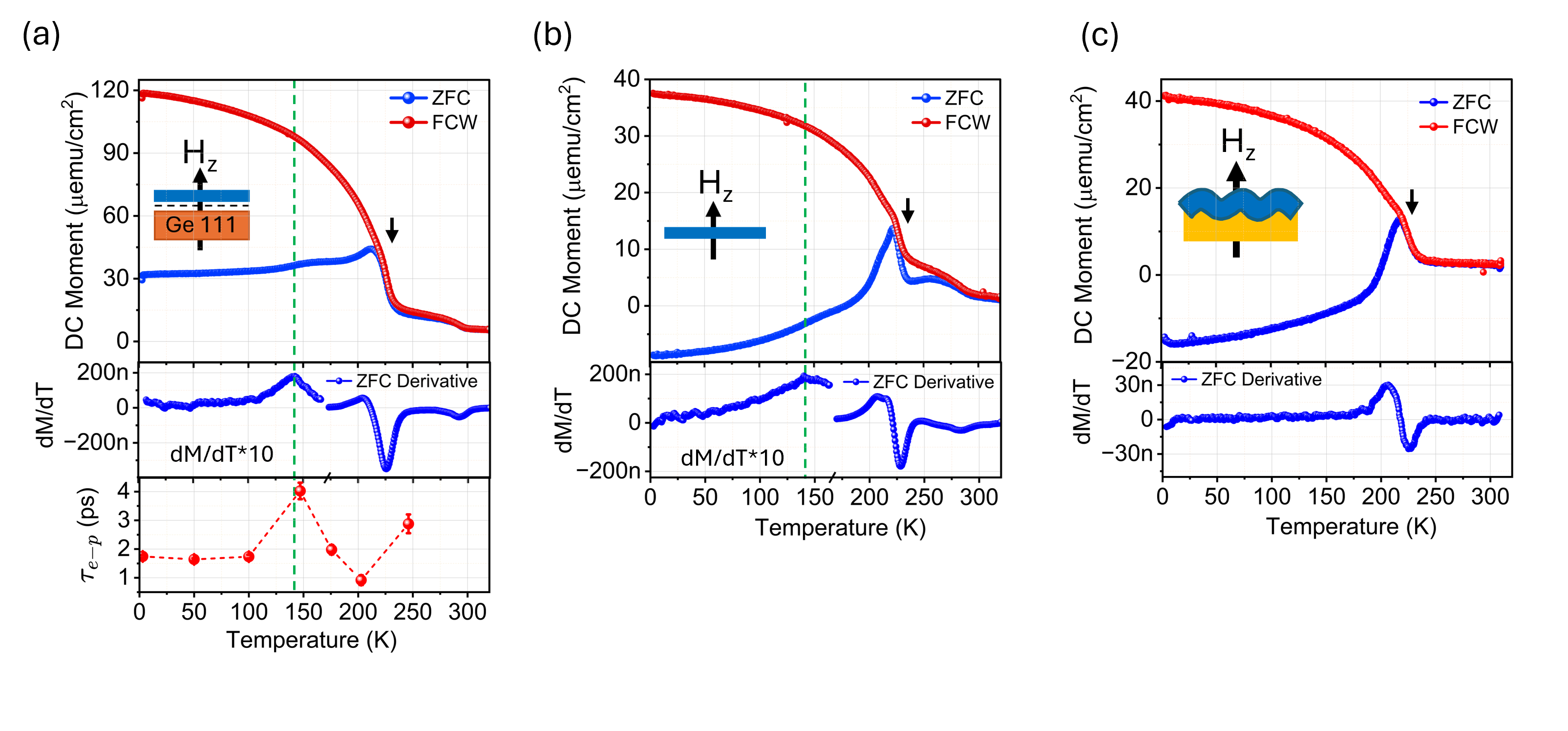}
\caption{\textbf{Persistent magnetic transitions and optical identification of putative SDW for MnPtGa film and released flat membrane, compared to rippled membrane.} (a) Superconducting quantum interference device (SQUID) magnetometry $M(T)$ and $dM/dT$ of MnPtGa/graphene/Ge(111) film together with the characteristic relaxation time extracted from time-resolved reflectivity.  (b) Released MnPtGa membrane. (c) Rippled membrane. All SQUID measurements are performed out of plane with $H_z=100$ Oe.
   }   
    \label{fig:magnetism}
\end{figure*}

\begin{figure*}[t]
    \centering
    \includegraphics[width=2\columnwidth]{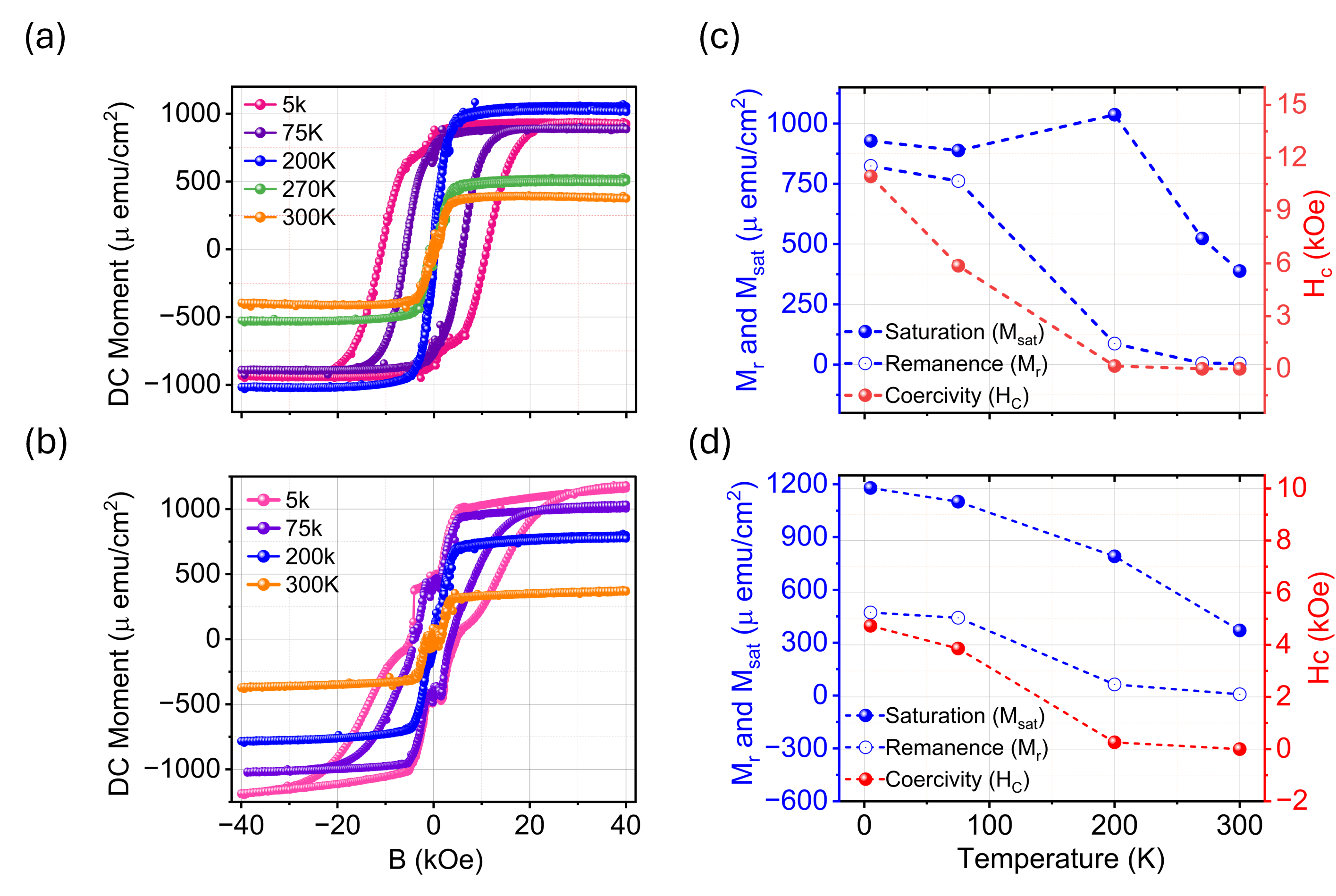}
    \caption{\textbf{Intermediate-temperature magnetic reorientation in supported MnPtGa film and released membrane.}
    (a) Isothermal magnetization versus applied field ($H \parallel z$) for the MnPtGa/graphene/Ge film and (b) released MnPtGa membrane. (c) Extracted parameters from $M(H)$ for the film and (d). For the film, the strong separation between $M_r$ and $M_{\rm sat}$ around 200 K suggests the onset of spin canting. In contrast, for the membrane the $M_r$ and $M_{\rm sat}$ both vary monotonically with temperature.}
    \label{fig:magnetism2}
\end{figure*}

SQUID magnetometry of both the film on graphene/Ge and released membrane reproduce the key features observed for bulk crystals (Fig. \ref{fig:magnetism}). Focusing first on the film, we observe a bifurcation in the zero field cooled (ZFC) - field cooled (FC) temperature-dependent magnetization near 225 K, consistent with a ferromagnetic (FM) onset. Further cooling produces a kink in the ZFC magnetization at 140 K, consistent with a spin reorientation, which is more clearly identified via a peak in the derivative $dM/dT$. 
A similar ZFC kink was previously assigned to a $k=0$ canted state in epitaxial MnPtGa films, whereas bulk MnPtGa develops a canted+SDW state below $\sim 140$ K. \cite{ibarra2022noncollinear,cooley2020evolution}. The magnetometry anomaly alone therefore does not uniquely distinguish these possibilities.

To resolve this question, we turn to optical time-resolved reflectivity, which is well suited to few-nanometer-thick films and membranes (Methods). The extracted decay time exhibits a pronounced maximum near the peak in $dM/dT$ (Fig. \ref{fig:magnetism}(a), see Supplementary Information for representative reflectivity traces and fits). Such slowing is consistent with a quasiparticle–phonon bottleneck associated with density-wave gap opening \cite{chia2006quasiparticle,meng2024density,pogrebna2014spectrally}. Although spin reorientation can also modify electronic scattering, an SDW naturally produces such a bottleneck through partial gapping and Fermi-surface reconstruction. The coincident magnetic and relaxation-time anomalies therefore favor an SDW-containing state over a purely orientational $k=0$ spin reorientation, similar to the state identified by neutron diffraction in bulk MnPtGa \cite{cooley2020evolution}.

The released MnPtGa membrane on tape retains the same qualitative magnetic features as the supported film, including the FM onset near 225 K and the \(dM/dT\) peak near 140 K associated with the putative SDW onset. As a first demonstration of intentional mechanical tuning, the \(\sim 140\) K peak is suppressed in a rippled MnPtGa membrane formed by transfer onto a polymer under compression [Fig.  \ref{fig:magnetism}(c)]. This indicates that the putative SDW-containing state, while robust to release, is destabilized by stronger imposed strain and bending.

Finally, we compare film and flat membrane magnetometry, focusing on the intermediate-temperature regime near 200 K where bulk MnPtGa develops canted-antiferromagnetic order before entering the lower-temperature SDW-containing state. For the film on graphene/Ge [Fig. \ref{fig:magnetism2}(a,c)], clear hysteresis develops below the collinear FM onset \(T_c \sim 225\) K. The saturation magnetization \(M_{\rm sat}\) increases on cooling into the ferromagnetic regime, reaches a maximum near 200 K, and then decreases, while the remanent magnetization \(M_r\) and coercive field \(H_c\) continue to increase. This divergence is qualitatively consistent with the onset of noncollinear spin reorientation: developing antiferromagnetic correlations can reduce the net field-polarized moment while increasing anisotropy and resistance to reversal. Upon further cooling through the \(\sim 140\) K anomaly identified in Fig. \ref{fig:magnetism}, \(M_{\rm sat}\) decreases more strongly, consistent with continued evolution toward the putative SDW-containing state.

The released membrane exhibits a more heterogeneous magnetic response near 200 K, evidenced by the weaker separation between \(M_r\) and \(M_{\rm sat}\) and by pinched \(M(H)\) loops [Fig. \ref{fig:magnetism2}(b,d)]. We attribute this broadened response to spatial variations in anisotropy and local magnetic order arising from weak, nonuniform strain in the membrane adhered to the tape. In contrast, the lower-temperature \(\sim 140\) K peak in \(dM/dT\) remains well defined after release [Fig. \ref{fig:magnetism}(b)], indicating that the putative SDW-containing state is robust to these small unintended strain variations. In contrast, for intentionally rippled MnPtGa membranes the $\sim 140$ K feature is suppressed (Fig. \ref{fig:magnetism}(c)), showing that the low-temperature state remains mechanically tunable.

\section{Discussion}

We demonstrated epitaxial synthesis and release of single-crystalline MnPtGa membranes from graphene/Ge(111). The supported films exhibit bulk-like lattice parameters, consistent with reduced epitaxial clamping across the graphene interface and providing a possible explanation for the more bulk-like low-temperature magnetic behavior compared with previous films on sapphire. Time-resolved reflectivity of the supported film reveals a pronounced relaxation time peak near 140 K, consistent with a quasiparticle–phonon bottleneck associated with a SDW gap. Together with the coincident $dM/dT$ peak, this provides evidence for an SDW-containing state and establishes an optical fingerprint compatible with the small magnetic volumes of ultrathin films and membranes, where neutron diffraction is challenging. 
The $\sim 140$ K magnetic anomaly is preserved after release but suppressed by intentional rippling, demonstrating that the putative SDW-containing state is robust to weak unintended strain yet tunable under stronger imposed deformation.
MnPtGa membranes therefore provide a platform for strain- and bending-controlled manipulation of competing magnetic orders, with time-resolved reflectivity offering a compatible probe of their evolution under mechanical tuning.




\section{Acknowledgments}

This work was primarily support by the NSF through the University of Wisconsin Materials Research Science and Engineering Center (DMR – 2309000) (MnPtGa synthesis by RH, ZL; TR-reflectivity by FF; SQUID by RH; supervision by JX and JKK). Additional synthesis by TS and JKK was supported by the U.S. Department of Energy, Office of Science, Basic Energy Sciences, under award no. DE-SC0023958. J.K.K. acknowledges support from the Gordon and Betty Moore Foundation (DOI:10.37807/GBMF13808) for data analysis. We acknowledge the use of facilities and instrumentation in the Wisconsin Center for Nanoscale Technology. This Center is partially supported by the Wisconsin Materials Research Science and Engineering Center (NSF DMR-2309000) and by the University of Wisconsin–Madison. Graphene synthesis (K.S.) was supported by the U.S. Department of Energy, Office of Science, Basic Energy Sciences under award no. DE-SC0016007.

\section{Methods}
\textbf{Molecular beam epitaxy growth of MnPtGa.}
MnPtGa films are grown using a room temperature seed and anneal approach to improve film wetting, following Ref. \cite{laduca2024cold}. Briefly, MnPtGa is deposited at room temperature on graphene/Ge resulting in a $\sim 5$ nm thick amorphous film observed by reflection high energy electron diffraction (RHEED). Annealing to $375 \degree$C crystallizes the film, as observed by a streaky RHEED pattern. Continued MnPtGa growth to a total thickness of 11 nm at $375 \degree$C retains the streaky RHEED pattern indicative of a smooth epitaxial film surface (Fig. \ref{fig:structure}(a) inset).
Fluxes from the Mn and Ga effusion cells, and from the Pt e-beam evaporator, are measured in-situ by quartz crystal microbalance and calibrated to absolute scale by Rutherford Backscattering Spectrometry (RBS) on separate samples.

\textbf{Time-resolved pump-probe reflectivity.}
Time-resolved pump–probe reflectivity (TR-R) measurements were performed using 
synchronized outputs from an 80 MHz femtosecond laser system (Chameleon Discovery 
NX, Coherent), with 1040 nm light used as the pump and 800 nm light as the probe. The pump and probe beams were incident nearly normal to the sample surface and focused onto the flake through a 20× objective.  The pump spot diameter at the sample was approximately $8~\mu\mathrm{m}$, and the pump fluence was $450~\mu\mathrm{J/cm^2}$. Temperature-dependent measurements were performed in a closed-cycle optical cryostat (OptiCool, Quantum Design). 

The transient-reflectivity traces were fitted using 
\begin{equation}
    R(t) = A e^{-t/\tau_{e-ph}}+B e^{-t/\tau_{ac}} cos(\omega_{ac}t) + C,
\end{equation}
where $A$ and $\tau_{e-ph}$ are the amplitude and characteristic relaxation time of the nonoscillatory response, respectively. The second term describes the damped coherent acoustic-phonon oscillation, with amplitude $B$, damping time $\tau_{ac}$, angular frequency $\omega_{ac}$, $C$ accounts for a long-lived background offset. 

\section{Supplemental Information}

Contains extended structural characterization (RHEED versus anneal sequence, x-ray rocking curve, and reciprocal space map) and raw time resolved reflectivity with fits to extract $\tau_{e-ph}$.

\bibliographystyle{apsrev}
\bibliography{ref}


\end{document}